# The Ratio Law of the Structure Evolution and Stability for Ti$_n$O$_m$ ($n$=3-18, $m$=1-2$n$) Clusters

Hongbo Du[a,b*], Yu Jia[b,c], Chunyao Niu[b], Kaige Hu[d], Haifeng Li[a], Lingmin Yu[e]

[a]*Department of Physics, School of Science, Xi'an Technological University, Xi'an, Shaanxi, 710032, China*

[b]*International Laboratory for Quantum Functional Materials, and School of Physics and Engineering, Zhengzhou University, Zhengzhou, Henan 450052, China*

[c]*Key Laboratory for Special Functional Materials of Ministry of Education, and School of Physics and Electronics, Henan University, Kaifeng, 475001, China*

[d]*School of Physics and Optoelectronic Engineering, Guangdong University of Technology, Guangzhou 510006, China*

[e]*School of Material and Chemical Engineering, Xi'an Technological University, Xi'an, Shaanxi, 710032, China*

**Abstract**

Most theoretical investigations about titanium oxide clusters focus on (TiO$_2$)$_n$. However, many Ti$_n$O$_m$ clusters with $m \neq 2n$ are produced experimentally. In this work, first-principles calculations are performed to probe the evolution of Ti$_n$O$_m$ clusters. Our investigations show that for $n$=3-11, there exist one relatively stable specie; while for $n$=12-18, there are two relatively stable species: Ti-rich and O-rich species. HOMO-LOMO calculations show that the gap can be tuned by changing the size and configurations of Ti$_n$O$_m$ clusters. Our investigation provides insights into the evolution of cluster-to-bulk process in titanium oxide.

**Introduction**

With the vast application prospects of nanoparticles, understanding of the atomic aggregation process, nanoparticles's configuration and stability has become an important goal for both experimental and theoretical researches.[1] New physical and chemical properties emerge when the size of a material becomes smaller and smaller, and even down to nanoscale. For example, the movement of electrons and holes in semiconductor nanomaterials is primarily governed by the well-known quantum confinement effect, and the transport properties related to phonon are largely affected by the size and geometry of the materials.[1-4] The surface-to-volume ratio increases dramatically as the size of a material decreases. The relatively large surface area owing to small particle size facilitates reaction/interaction between the devices and the interacting media. Powders and dispersed catalysts are indeed more reactive than terraces of crystalline materials based on experiments.[5] Atomic clusters have long been considered as models for fundamental mechanistic insight into complex surfaces and catalysts. Illustrating the evolution behavior from the molecular to bulk by increasing the cluster size is very important for understanding nanotechnology principles and applications, especially for material selection and design.

Titanium dioxide ($TiO_2$) is an important material for various applications including electronic devices, polluting compounds decomposition, medical bio-engineering, solar energy conversion and water dissociation. Owing to its technological importance, numerous experimental and theoretical investigations have been carried out on isolated titanium oxide clusters to correlate their structures and properties with their bulk counterparts.[6-18] Yu and Freas produced isolated positive titanium/oxygen clusters by sputtering titanium dioxide powder and titanium foil (exposed to oxygen) with an energetic xenon atom beam (~8 keV).[7] The form of titanium/oxygen cluster ions from their results is $[Ti_nO_{2n-\delta}]^+$, where $n$ equals 1 to 8 and $\delta$ ranges from 0 to 4. Matsuda and Bernstein prepared neutral titanium oxide clusters ($Ti_mO_n$, $m$=1-7, $n$=1-12) in a supersonic expansion by laser ablation of the metal and reaction with oxygen in He expansion gas.[13] Weichman *et al.* presented the vibrationally resolved spectra of titanium dioxide cluster anions $(TiO_2)^{-n}$ for $n = 3-8$, using infrared photodissociation spectroscopy.[14] Liu *et al.* investigated bandgap engineering on titanium oxide clusters by substituting labile surface sites with ligand.[19] Based on their results, 14 O-donor ligands have been successfully introduced without changing the cluster core. They reported a powerful approach to adjust the bandgap of titanium oxide clusters, and also provided a perfect model library for related theoretical studies.

At present, theoretical studies on titanium oxide mainly aimed at $(TiO_2)_n$ ($n\leq24$)[8, 9, 11, 12, 14-16] and experimental researches focused on $Ti_nO_m$ ($n$=1-7, $m$=2$n$-3, 2$m$+1)[7, 13, 14]. A clear, systemic and an in-depth understanding for the geometries of $Ti_nO_m$ particles have not yet emerged. In this paper, our computational results show that besides $(TiO_2)_n$, some cluster species with an O/Ti ratio $m/n<2$ can also be stable, in consistent with earlier experiments.[7, 13] First-principles calculations have been performed to probe the evolution process of stable structures of $Ti_nO_m$ ($n\leq18$, $m\leq2n$). Our work is important to understand the structure evolution of titanium oxide from small clusters to bulk.

## Method

Calculations based on density functional theory (DFT) were carried out using the Vienna *ab initio* simulation package (VASP), [20, 21] with the exchange-correlation energy corrections described by the Perdew–Burke–Ernzerh parametrization.[22] The interactions between the valence electrons and the ionic cores were described by the projector augmented wave method, [23, 24] the geometric structures were optimized by conjugated gradient (CG) method, [25] and the wave function was expanded in a plane wave basis with the energy cutoff of 400 eV. The simple cubic unit cells are used, and a vacuum region of more than 12 Å was applied in order to minimize image interactions. The total energy is converged to $10^{-4}$ eV for the electronic structure relaxations and the convergence criterion for the force on each ion is 0.02 eV/Å.

The initial structures of $Ti_nO_m$ ($n\leq18$, $m\leq2n$) used in our calculations were obtained from our rational design. We used the Monte Carlo method to generate a large number of initial structures for each size of clusters, following a scheme of a stochastic search for isomers on a quantum mechanical surface by Saunders in which a minimum-energy structure is subject to a motion to each atom with a random distance within a maximum distance in a random direction.[26] We generated about 50-100 initial structures for $Ti_nO_m$ when $n \leq 6$ and more than 100 initial structures when $n>6$. We found that although we have chosen a large number of initial structures, most of them reach only a few configurations after relaxation. Next, as we known, the number of possible configurations of clusters increases dramatically as the increase of the cluster size. It is difficult to create all possible initial structures. For smaller ones, it is relative easy to obtained possible stable structures by test all the possible non-degenerate state geometries. From the stable structures of smaller clusters ($n \leq 8$) we obtained, we found a general rule: Ti atoms of stable geometries of $Ti_nO_m$ favor to have three-fold or four-fold coordination. In addition, as we know that larger clusters prefer to have a smaller surface-to-volume ratio. So based on this rule, we further constructed new possible structures that were not included in the initial random configurations. We also used another complementary method, which has been adopted in literature and turns out to be useful in finding new structures.[27-29] This method constructs new initial structures based on the previously determined geometries of smaller clusters in which a certain amount of atoms was added; an inverse routine can also be adopted to construct alternative initial structures in which a certain amount of atoms was removed from the previously determined geometries of lager clusters. At last, molecular dynamical method was used to ascertain the stability of the clusters.

## Results and Discussion

Based on the above methods and rules, we obtained the lowest-energy structures of $Ti_nO_m$ ($n$=3-11, $1 \leq m \leq 2n$; $n$=12-18, $n \leq m \leq 2n$), whose stability has been carefully checked. Binding energies of the lowest-energy structures of $Ti_nO_m$ ($n$=3-18, $m \leq 2n$) from DFT calculations are shown in Fig. 1. The relative stability of $Ti_nO_m$ clusters as a function of the O/Ti ratio was investigated via the analysis of binding energies per oxygen atom $E_b$, defined with respect to Ti atom and 1/2 $O_2$:

$E_b = (E_{Ti_nO_m} - nE_{Ti} - (m/2)E_{O_2})/m$. For Ti$_n$O$_m$ clusters with $n$ equals to 3-6, we optimized a wide range of oxygen ratios, $m=1$-$2n$. Figure 1 shows that the relatively more stable clusters of Ti$_n$O$_m$ ($n=3$-6) according to our rule are Ti$_3$O$_5$, Ti$_4$O$_6$, Ti$_5$O$_8$ and Ti$_6$O$_9$. It means the probability of existence of these species are higher than other species in experiments. All the value of $m/n$ for the relatively stable structures are larger than 1, which is consistent with earlier reports.[7, 13] In addition, the clusters of $m<n$ are relatively not stable. Considering this feature, for larger clusters Ti$_n$O$_m$ ($n=7$-18), we only calculated the cases of $n \leq m \leq 2n$. The subfigures in Fig. 1 for each fixed $n$ of Ti$_n$O$_m$ with $n=3$-11 shows only one valley in each curve, indicating there is one relatively most stable specie when $m$ varies. However, two valleys emerge when $n=12$-18, which means there are two relatively stable species for $n=12$-18. We define the two relatively stable species of Ti$_n$O$_m$ ($n=12$-18) as Ti-rich species whose O/Ti ratios are in the range of 1.14-1.44 and O-rich species whose O/Ti ratios are in the range of 1.64-1.875. For each n, there are two species whose probability of existence are higher than other species. Earlier published results proved that different species exist for each $n$ of small Ti$_n$O$_m$ clusters with different probability which is consistent with our results.

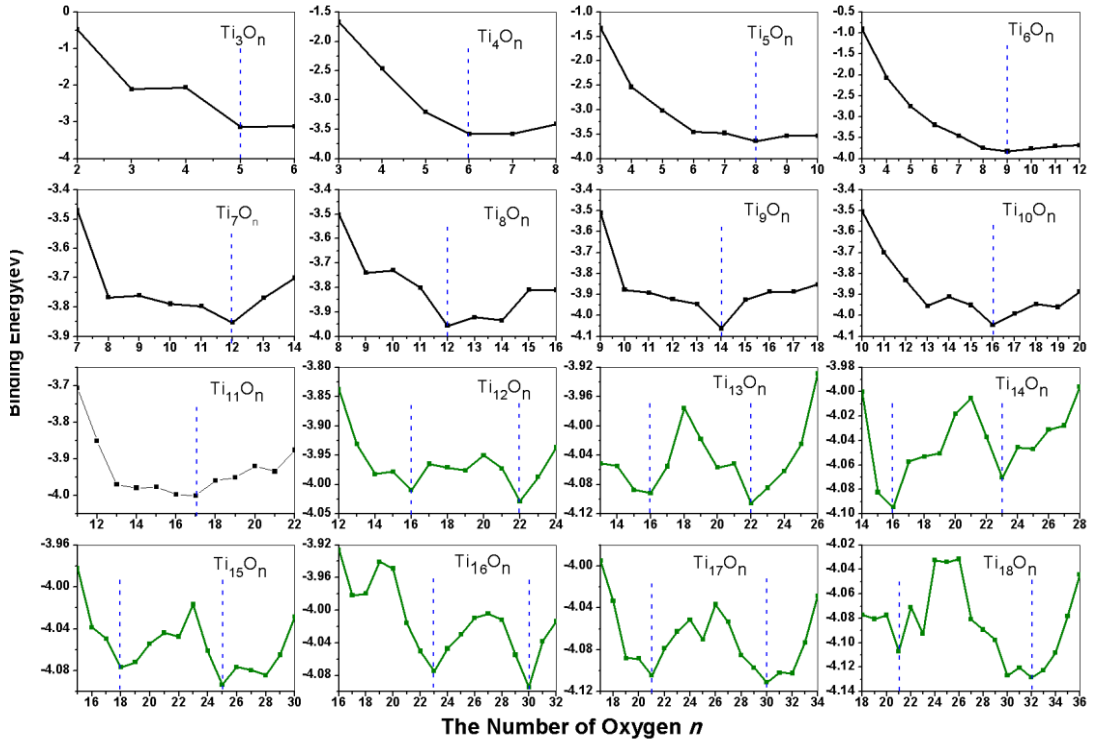

Figure 1. Binding energies of the lowest-energy structures of Ti$_n$O$_m$ ($n=3$-18, $m \leq 2n$) as function of the number of oxygen based on DFT calculations.

Figure 2 plots $m$ versus $n$ as for relatively stable species of $Ti_nO_m$ ($n$=3-18). The O/Ti ratios of stable species of $Ti_nO_m$ ($n$=3-11) are between 1.50 and 1.71, the values of Ti-rich species $Ti_nO_m$ ($n$=12-18) are between 1.14 and 1.44, and the values of O-rich species $Ti_nO_m$ ($n$=12-18) are between 1.64 and 1.875. As for $n$=12-18, two sets of relatively more stable species exist.

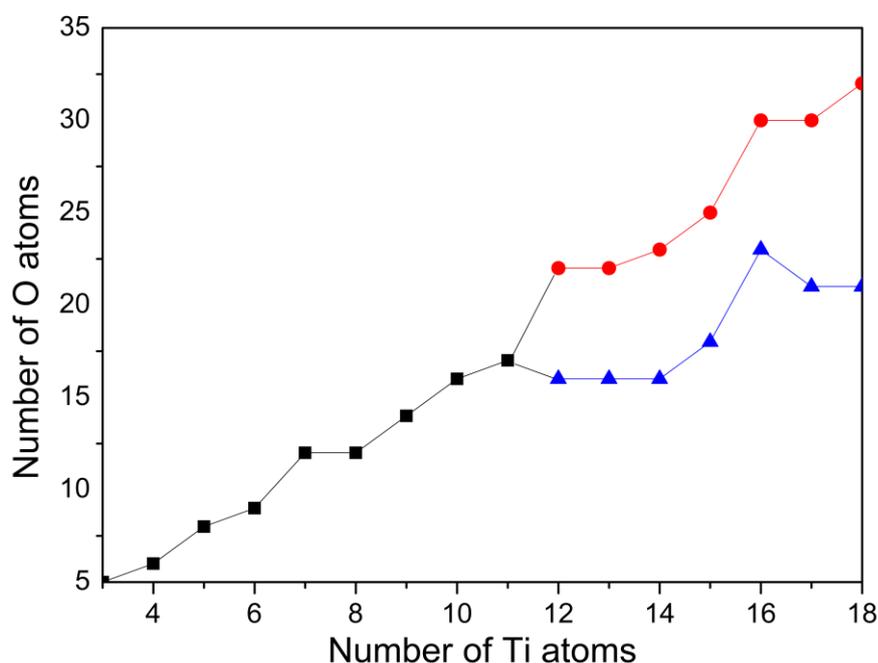

Figure 2. The number of oxygen atoms $m$ as function of the number of titanium atoms $n$ for stable species of $Ti_nO_m$ ($n$=3-18).

Figure 3 shows the geometries of stable species of $Ti_nO_m$ ($n$=3-11) and also the most stable structures of $(TiO_2)_n$ ($n$=3-11). It can be seen that all the Ti atoms are on the surface of the clusters except $Ti_{11}O_{17}$. Ti atoms in the relatively stable species of $Ti_nO_m$ ($n$=3-11) bond with three oxygen atoms are similar to the bond configuration of ammonia, and those bond with four oxygen atoms are similar to the bond configuration of methane. Terminal oxygen atoms are rare in the relatively stable structures of $Ti_nO_m$ ($n$=3-11), while terminal oxygen atoms commonly exist in the geometries of lowest-energy isomers of $(TiO_2)_n$ ($n$=3-11). The relatively stable structures of $Ti_nO_m$ ($n$=3-11) prefer to have three-fold coordination and four-fold coordination Ti atom and it is energy penalty for the dangling Ti-O bonds.

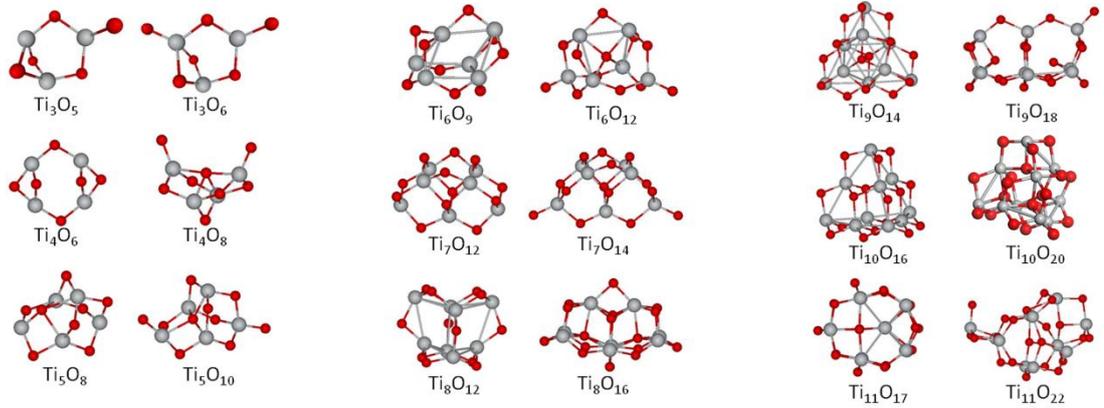

Figure 3. The relatively more stable species for $Ti_nO_m$ ($n$=3-11) and the most stable structures of $(TiO_2)_n$ ($n$=3-11). The gray spheres represent titanium atoms and the red spheres represent oxygen atoms.

Figure 4 shows the two relatively stable geometries for each $Ti_nO_m$ with $n$=12-18, i.e., Ti-rich stable structures and O-rich stable structures, displayed as the first and second columns, respectively. The lowest-energy structure of $(TiO_2)_n$ ($n$=12-18) are also displayed, shown as the third column. It turns out that Ti-rich stable structures are more spherical and compact relative to O-rich stable structures and $(TiO_2)_n$ structures. For the Ti-rich stable structures, a central Ti core can be found bonding to surface Ti atoms, at the same time the surface Ti atoms prefer to bond to three oxygen atoms. The number of central oxygen atoms increases with $n$ because of the expansion of the central space in clusters. This suggests only the surface Ti atoms are oxidized by oxygen atoms for Ti-rich structures. For O-rich stable structures, the Ti atoms are four-fold coordinated for $Ti_nO_m$ ($n$=12,13,16,18), but for $Ti_nO_m$ ($n$=14,15, 17), most Ti atoms at the surface are three-fold coordinated and with a Ti core in the center. The oxidation process are deep into the cluster centers. Now we turn to the stable geometries of $(TiO_2)_n$ ($n$=12-18). In fact, many works have been done on the stable structures of $(TiO_2)_n$ ($n$=2-24) and most geometries listed in Figs. 3 and 4 can be found in literature , except $(TiO_2)_{15}$ , whose energy is 2.0 eV lower than earlier reports [8, 9, 12, 14, 15]. In the third column of Fig. 4, We can see that there is no dangling Ti-O bond on the surface of $(TiO_2)_n$ ($n$=12-18). The above results suggest that the oxidation pattern of the small titanium clusters is from surface to central.

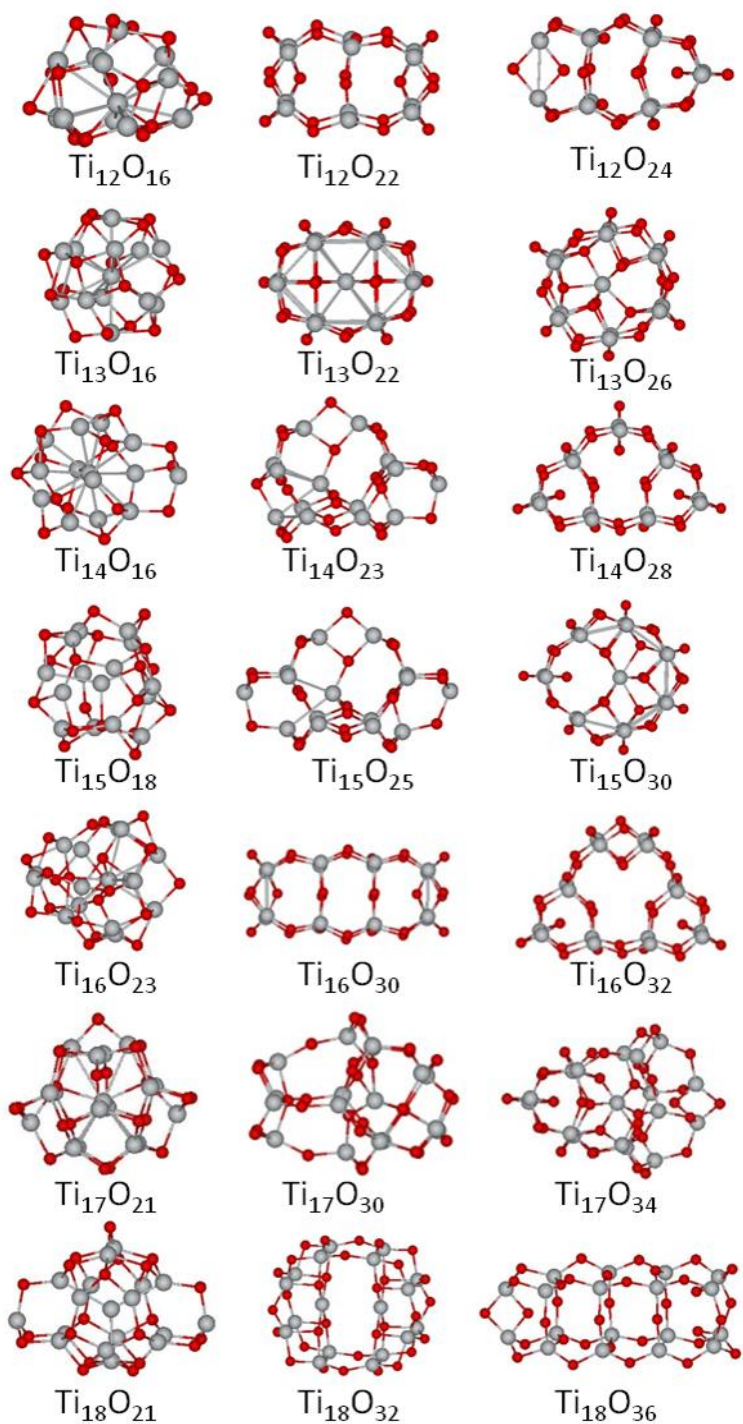

Figure 4. The relatively more stable species for $Ti_nO_m$ ($n$=12-18) and the most stable structures of $(TiO_2)_n$ ($n$=12-18). The gray spheres represent titanium atoms and the red spheres represent oxygen atoms.

Among all the stable geometries shown in Figs. 3 and 4, $Ti_{11}O_n$ are transition species, which are obviously important to understand the evolution mechanism from small clusters to bulk. More oxygen atoms exist on the surface of O-rich species means a larger center space to contain more atoms. From Fig. 1, we can see that the differences of stability among stable $Ti_{11}O_{15}$, $Ti_{11}O_{16}$, $Ti_{11}O_{17}$ are small. It suggests the differences of probabilities of existence for $Ti_{11}O_{15}$, $Ti_{11}O_{16}$, $Ti_{11}O_{17}$ are small. Figure 5 displays the lowest-energy structures of $Ti_{11}O_{15}$, $Ti_{11}O_{16}$, $Ti_{11}O_{17}$ and corresponding structures whose energy difference relative to lowest-energy structures are less than 0.2 eV. It turns out that the relatively stable structures of $Ti_{11}O_{15}$ is hemispherical, similar to the Ti-rich structures of $Ti_{12}O_{16}$ in Fig. 4. The common hemispherical shape of $Ti_{11}O_{15}$ and $Ti_{12}O_{16}$ because there are not enough atoms to form closed surface around the Ti core. For $Ti_{11}O_{16}$, one Ti atom and one O atom exist in the center of the two structures and all the surface O atoms bond with two Ti atoms, which means the surface area as well as the central space volume reach their maximums. For $Ti_{11}O_{17}$, two oxygen atoms and one Ti atom locate in the cluster center. $Ti_{12}O_{16}$, $Ti_{11}O_{17}$ and $Ti_{12}O_{22}$ have similar feature that all the surface O are two-fold coordinate and all the Ti atoms are oxidized.

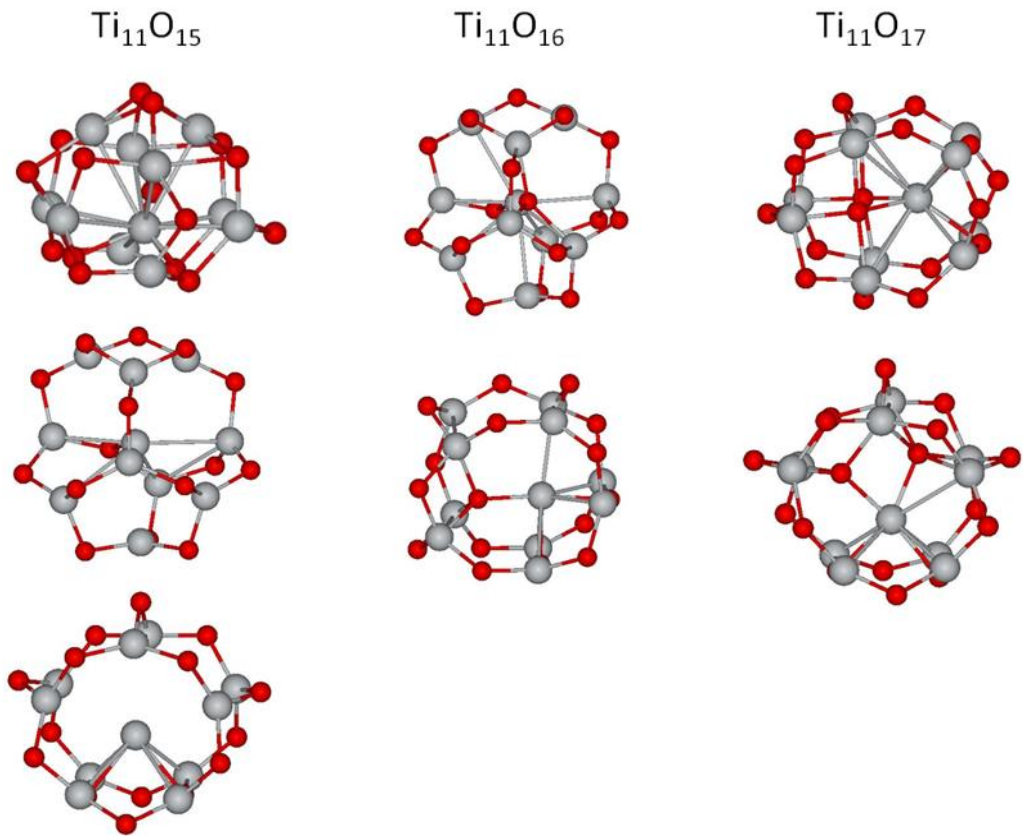

Figure 5. Lowest-energy structures of $Ti_{11}O_{15}$, $Ti_{11}O_{16}$ , $Ti_{11}O_{17}$ and corresponding

energy degenerated structures which their energy difference relative to lowest-energy structures are less than 0.2eV. The gray spheres represent titanium atoms and the red spheres represent oxygen atoms.

Figure 6 shows the HOMO-LOMO gap of $Ti_nO_m$, where $m$ is taken as a variable for each given $n$ ($n$=3-18). The HOMO-LOMO gaps of $(TiO_2)_n$ are between 1.5 eV and 3.0 eV, clearly higher than other species, whose values are between 0.1 eV to 1.5 eV. These results shows that HOMO-LOMO gap can be tuned by changing the size and structures of nanoclusters, providing a way to further tune their electronic and optical characters.

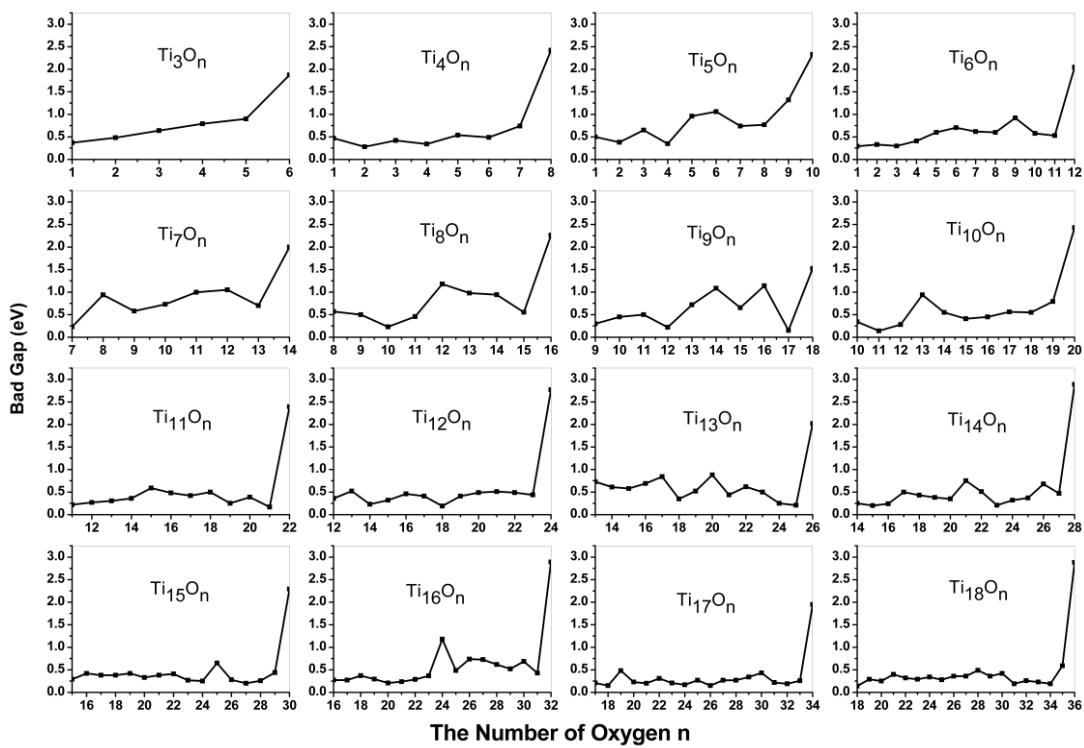

Figure 6. HOMO-LOMO gap of $Ti_nO_m$ as the number of $m$ when $n$ is fixed ($n$=3-18).

Conclusion

The evolution of lowest-energy structures of $Ti_nO_m$ ($n$=3-11, 1≤$m$≤2$n$; $n$=12-18, $n$≤$m$≤2$n$) clusters have been studied by first-principles calculations. Our investigation suggests that the ratio of $m/n$ of relatively stable species is in the range of 1.5-1.71 for $n$=3-11. Two relatively stable species of $Ti_nO_m$ can be found for $n$=12-18: Ti-rich species with the O/Ti ratios in the range of 1.14-1.44 and O-rich species with the O/Ti ratio in the range of 1.64-1.875. It means the probability of existence of these species are higher than other species in experiments. Our results indicate that the oxidation pattern of small titanium clusters is from surface to center. In addition, we found that the HOMO-LOMO gaps of $Ti_nO_m$ decrease as $n$ increases. The values of HOMO-LOMO gaps of $Ti_nO_m$ are in the range of 0.1-1.5 eV. Our results suggest that the HOMO-LOMO gap can be tuned by changing the cluster size. The evolution of stable structures of $Ti_nO_m$ ($n=3-11$, $1 \leq m \leq 2n$; $n=12-18$, $n \leq m \leq 2n$) clusters may provide useful insights into the evolution of cluster-to-bulk process in titanium oxide.

The clusters studied in this work are small in size (less than 1 nm) relative to experimental crystalline titanium oxide clusters (usually larger than 5nm). Most atoms locate at the cluster surface whose sizes are less than 1 nm. As a result, they don't have experimental periodic crystalline structure. As the size of titanium oxide clusters gets larger, more atoms locate at the center of their structures, then transitional property of structures should appear. And this should be demonstrated in further work.

**Acknowledgements**
The work was supported by Natural Science Foundation of Shaanxi Province Department of Education (16JK1384，17JK0374).